# A novel method for determining the resistivity of compressed superconducting materials

Liling Sun[1,2]†, Qi Wu[2], Shu Cai[1], Yang Ding[1] and Ho-kwang Mao[1]

[1]*Center for High Pressure Science & Technology Advanced Research, 100094 Beijing, ChinaCenter*
[2]*Institute of Physics, Chinese Academy of Sciences, Beijing 100190, China*

The resistivity of a superconductor in its normal state plays a critical role in determining its superconducting ground state [1,2]. However, measuring the resistivity of a material under high pressure has long presented a significant technical challenge due to pressure-induced changes in the crystallographic directions, especially for samples with anisotropic layered structures like high-$T_c$ superconductors and other intriguing quantum materials. Here, we are the first to propose a novel and effective method for determining high-pressure resistivity, which relies on the ambient-pressure resistivity, initial sample sizes, lattice parameters, high-pressure resistance, and lattice parameters measured from the same sample. Its validity has been confirmed through our investigations of pressurized copper-oxide superconductors[3], which demonstrates that this method provides new possibilities for researchers conducting high-pressure studies related to resistivity of these materials.

**Confronted problem**

Apart from chemical doping, pressure is the most important tuning method. Applying pressure to superconductors or other related quantum materials not only unveils numerous novel physical phenomena [4-17], providing vital insights for the exploration of new superconductors[18,19], but also can effectively assists in uncovering the associated mechanisms [20,21].

Electrical resistivity is a fundamental characterization of superconducting and numerous quantum materials. It is established that the resistivity ($\rho$) can be described by the following equation:

$$\rho = 1/\sigma = R \times (S/L) \qquad (1)$$

where σ represents the conductivity, $R$ stands for the resistance, $L$ denotes the distance between the electrical leads for the voltage measurements, $S$ represents the cross-section of the sample. In practical terms, determining the resistivity of a compressed material in a diamond anvil cell (DAC) is challenging due to the high-pressure experimental setup, which makes direct measurement of the pressure-induced size change of the sample in the three crystallographic directions difficult (see Fig. 1), especially for materials with anisotropic layered crystal structures, such as high-temperature superconductors and other intriguing materials. Consequently, people have had to use resistance, an extensive quantity, to describe the conducting property of the compressed materials, which hinders the identification of universal trends through comparison with earlier experimental results obtained from ambient-pressure chemical doping. An important example that underscores the significance of our proposed method is the investigation of the correlation among superconducting transition temperature ($T_c$), the superfluid density ($\rho_{sc}$), and the normal state conductivity (σ) at $T_c$ (*i.e.*, σ is the reciprocal of resistivity), which is a crucial aspect in understanding "why $T_c$ is high" in high-$T_c$ materials[2], through the application of the Homes' law that establishes a scaling relation of $T_c$ with $\rho_{sc}$ and resistivity (ρ) ($\rho_{sc} = AT_c/\rho$)[1,2]. Therefore, the development of an adoptable method to determine high-pressure resistivity is crucial for the contemporary high-pressure quantitate studies on numerous enigmatic quantum materials.

**Proposed method**

First, let us consider the ambient-pressure case (see Fig.1 and equation (1)), where the distance between the electrical leads for voltage measurements ($L$), the cross-section ($S$) and resistance ($R_0$) of the ambient-pressure sample can be directly measured, allowing the determination of the ambient-pressure resistivity $\rho_0$. Given that $S = na_0 \times mc_0$ and $L = lb_0$, $\rho_0$ can be represented as:

$$\rho_0 = R_0 \times [(na_0 \times mc_0)/lb_0] = R_0(nm/l)(a_0c_0/b_0) \qquad (2)$$

Here $a_0$, $b_0$ and $c_0$ represent the ambient-pressure lattice parameters of the sample's unit cell, while $n$, $m$ and $l$ denote the numbers of the unit cell of the measured sample in three crystallographic directions. The value of *nm/l* can be calculated based on the initial size of the loaded sample. As the pressure is applied, $a_0$, $b_0$ and $c_0$ change with increasing pressure, but *nm/l* remains constant in the absence of a crystal structure phase transition. Therefore, the relationship between the resistivity ($\rho_i$) and the measurable resistance ($R_i$) at the fixed pressure ($P_i$) can be expressed as:

$$\rho_i = R_i \, (nm/l) \, (a_i c_i / b_i) \qquad (3)$$

Here, $a_i$, $b_i$ and $c_i$ denote the lattice parameters at $P_i$, which can be obtained from the high-pressure x-ray diffraction measurements. By employing equation (3), we can derive $\rho_i$ at different pressures.

**The requirements for adopting this method**

It is important to note that when employing this method, the sample must meet the following criteria:

1. The measured sample should be a bulk single crystal, and its ambient-pressure sizes (*S, L*), lattice parameters ($a_0$, $b_0$, $c_0$) and resistivity, as well as its high-pressure lattice parameters ($a_i$, $b_i$, $c_i$) and the resistance ($R_i$) should be known.
2. The samples with cubic, tetragonal and orthorhombic structures are suitable for obtaining more accurate $\rho_i(P_i)$.
3. No pressure-induced structural phase transition should occur within the pressure range investigated.
4. When studying the low-temperature properties of materials, using low-temperature x-ray diffraction (XRD) to measure the lattice parameters at both ambient pressure and high pressure, instead of room temperature XRD, should yield more precise results. This is because the values of $a_0$, $b_0$ and $c_0$ as well as $a_i$, $b_i$ and $c_i$ collected at low temperature may differ to some extent from those measured at room temperature.

**Significance of the method**

This method provides a viable solution to overcome the technical challenges associated with determining high-pressure resistivity. We have validated the effectiveness of this approach by examining the pressure-induced coevolution of superconductivity with resistivity and superfluid density in bismuth-based cuprate superconductors[3].

The scaling results demonstrate that this method is applicable to these high-$T_c$ superconducting materials. Moreover, our results show that it not only introduces researchers to a new means of obtaining high-pressure resistivity but also contributes to understanding the underlying superconducting mechanism by generating a substantial amount of high-pressure data regarding the correlation of superconducting transition temperature with superfluid density and resistivity on the scaling line of the universal Homes' law. Additionally, it is anticipated that this method can also be utilized in high-pressure studies on other significant materials, thereby advancing the field of the contemporary condensed matter physics and material science.

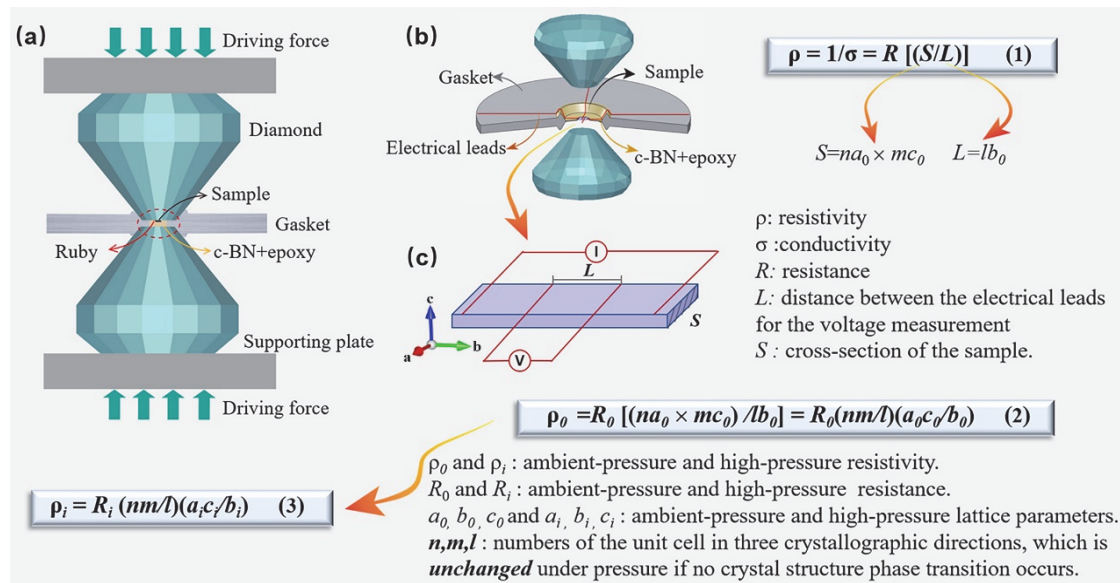

Figure 1 (a) A schematic of the high-pressure experimental setup for resistance measurements in a diamond anvil cell, where a mixture of c-BN and epoxy serves as an insulating material and ruby is utilized as a pressure calibrator for determining the sample pressure. (b) An enlarged view illustrating the arrangements of the sample in the gasket hole. A standard four-probe method is used to establish contact with the sample, while the insulating material is employed to isolate it from the metallic gasket. (c) The measured sample and the technique for resistance measurement. The three equations provide an explanation of the method for determining the resistivity of compressed materials.

**Acknowledgements**

This work was supported by the National Key Research and Development Program of China (Grant No. 2022YFA1403900 and 2021YFA1401800) and the NSF of China (Grant Numbers Grants No. U2032214, 12104487).

**Author information**

The authors declare no competing financial interest. Correspondence and requests for materials should be addressed to L.S.(liling.sun@hpstar.ac.cn or llsun@iphy.ac.cn).

**Data availability**

The data that support the findings of this study are available from the corresponding author upon reasonable request.